\documentstyle[multicol,epsfig,aps,prl,array]{revtex}

\newcommand{\htl}{{\tilde{h}}}

\newcommand{\rhobar}{{\overline{\rho}}}
\begin{document}
\topmargin0.cm
\title{Roughness of Sandpile Surfaces}
\author{J. G. Oliveira$^1$, J. F. F. Mendes$^1$ and
G. Tripathy$^2$ } \address{$^1$ Departamento de F\'{\i}sica,
Universidade de Aveiro, Campus Universit\'ario de Santiago, 3810-193
Aveiro, Portugal \\ $^2$ Institute of Physics, Sachivalay Marg,
Bhubaneswar 751005, India}
\date{15 Dec 03: ver14.tex} \maketitle

\begin{abstract}
We study the surface roughness of prototype models displaying
 self-organized criticality (SOC) and their noncritical variants in
 one dimension.  For SOC systems, we find that two seemingly
 equivalent definitions of surface roughness yields different
 asymptotic scaling exponents.  Using approximate analytical arguments
 and extensive numerical studies we conclude that this ambiguity is
 due to the special scaling properties of the nonlinear steady state
 surface. We also find that there is no such ambiguity for non-SOC
 models, although there may be intermediate crossovers to different
 roughness values. Such crossovers need to be distinguished from the
 true asymptotic behaviour, as in the case of a noncritical disordered
 sandpile model studied in \cite{barker00}.

\end{abstract}
\pacs{PACS numbers: 5.40-a, 5.70.Ln, 61.50.Cj}

\begin{multicols}{2}

Since the original proposal by Bak, Tang and Wisenfeld \cite{btw}
there has been a large body of work directed towards understanding
ubiquity of scale invariance in externally driven open dissipative
systems using the concept of self-organized criticality (SOC). The
sandpile is a prototype model system which has been extensively used
as a paradigm of SOC \cite{SOCreview,kadanoff}.  The principal aim has
been to elucidate how slowly driven dissipative systems with fast
relaxation mechanisms display long-tailed distributions of activity
sizes.  Recently, there have been resurge of interest in sandpile
models in order to understand SOC in connection with better understood
scale invariance in other nonequilibrium phenomena such as absorbing
state phase transitions \cite{soc&dp} and driven interfaces \cite{soc&sg}.

In \cite{krug92}, Krug, Socolar and Grinstein (KSG) studied the
surface fluctuations in a prototype model of SOC, the {\it limited
local sandpile} (LLS) \cite{kadanoff} and its variations in one spatial
dimension and concluded that the interfacial fluctuations, although
nontrivial, are evidently unconnected to the criticality of the
system. In fact, they argued that the evolution of height fluctuations
$\htl(x,t)$ can be described by an extension of the Kardar-Parisi-Zhang
(KPZ) \cite{kpz,krug97} equation for an anchored interface
\begin{equation}
\partial_t \htl=D\partial_x^2\htl+ c\partial_x\htl+\lambda
(\partial_x\htl)^2+\eta(x,t)\,,
\label{eq:htevol}
\end{equation}
where $\eta(x,t)$ is a Gaussian white noise. The linear term
$c\partial_x\htl$ is more relevant than the diffusion term
$D\partial_x^2\htl$ and the KPZ nonlinearity $\lambda
(\partial_x\htl)^2$ (in RG sense) and, since the interface is anchored, 
can not be eliminated by a Galilean shift. This
term is responsible for transporting fluctuations up the pile thus
relating the spatial roughness of the anchored interface to the
dynamical roughening of a moving KPZ interface in $d=1$,
$\alpha_{LLS}=\beta_{KPZ(1d)}=1/3$. This argument can be easily
extended to higher dimensions.
\begin{figure}[tb]
    \centering \includegraphics[width=\linewidth]{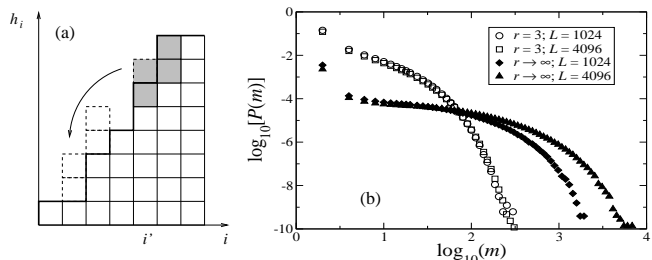}
    \caption{(a) Dynamics of the sandpile: An initial surface (thick
    line) on becoming unstable relaxes by transferring the unstable
    cluster (shaded blocks) successively to the left till a stable
    configuration is reached. (b) Log--log plot of the avalanche size
    distribution $P(m)$ for the LLS (solid symbols) and the ILLS (open
    symbols) for two different system sizes.  $P(m)$ for the ILLS
    falls off with a characteristic length scale independent of $L$. }
\label{fig:sandpile}
\end{figure}
In this paper we study interface roughness in LLS and related models
and find that effect of the criticality of the system is somewhat
subtle and is due to the nontrivial nature of the steady state
surface. The value of surface roughness depends on whether it is
measured about this nontrivial steady state profile or about the mean
instantaneous surface, the latter being the correct choice for moving
interfaces. The assertion made by KSG is valid only if $\htl$ in
(\ref{eq:htevol}) is defined as the fluctuation around the nonlinear
steady state surface. We also study modifications of the LLS which
have nonlinear steady state profiles but which do not display SOC and
find that there is no such ambiguity in the asymptotic roughness
exponent.  Thus, we conjecture that the roughness exponent is uniquely
defined for systems not displaying SOC even though they may possess
nontrivial steady state surfaces \cite{foot0}. For SOC systems the
ambiguity exists and one needs to define the roughness appropriately.
Our results have important bearings on studies such as the recent one
by Barker and Mehta \cite{barker00} who observed anomalously large
roughness exponent in a sandpile model with structural disorder.

The limited local sandpile model in $d=1$ is defined as follows (see
Fig.~\ref{fig:sandpile}a): on a one dimensional lattice of sites
$i=0,1,\cdots,L$ we define an integer height variable $h_i$. One grain
of sand added at a randomly chosen site increases the height at that
site by one: $h_i\rightarrow h_i+1$. The configuration is stable if
the local slopes satisfy $z_i = h_{i+1}-h_{i} \leq z_c$ for all $i$,
where we chose $z_c=2$ (our results are essentially unchanged
for $z_c\ge 2 $). An instability occurs when by addition of
grains at site $i+1$, the local slope exceeds threshold $z_i>z_c$; in
this case $z_c$ grains are transferred from column $i+1$ to column
$i$. Subsequently, columns $i$ and $i+2$ may become unstable, setting
off further topplings leading to an avalanche. The grains involved in
an avalanche leave the system if they reach site $i=0$ ($h_0(t)=0,
\forall t$). A new grain is added after the system has attained a
stable configuration. The final configuration reached is independent
of the order in which the sites are updated in case more than one site
become unstable. We count one unit of time (Monte Carlo step) for
every $L$ grains added.

The standard way of quantifying avalanches in these systems is to count
the number $m$ of grains that leave the pile after each deposition
\cite{kadanoff}. The criticality of the model is reflected in a broad
probability distribution of avalanche sizes, $P(m)$
(Fig.~\ref{fig:sandpile}b), which has no other length scale except
the system size $L$.

In \cite{krug92}, KSG also introduced and studied the {\it inertial}
limited local sandpile (ILLS) which mimics the effect of inertia of
the falling grains in a real sand pile. The instability condition
setting off an avalanche is same as that in the LLS but the condition
of stopping is changed. A cluster of grains when first destabilized is
assigned a momentum $p=0$ and each time the front of the cluster
reaches an unstable site ($z_i>z_c$) it gathers momentum,
$p\rightarrow p+1$. If it comes across a stable site, $p$ decreases by
an amount $r$: $p\rightarrow p-r$. The cluster continues moving down
as long as $p>0$ and leaves the pile if it reaches the site
$i=0$. Clearly, $r=\infty$ corresponds to the LLS. It was noted by KSG
that ILLS is not critical for any finite value of $r$ which is
reflected in the corresponding avalanche size distribution
(Fig.~\ref{fig:sandpile}b).

To study the interfacial fluctuations in these models, we start with
an initially flat pile ($h_i(0)=0, ~\forall i$) and add grains till
the pile reaches a steady state with the mean surface making a
critical angle $\psi$ ($\tan\psi_{L\rightarrow \infty}=3/2$) with the
horizontal. The closed boundary condition at $i=L$ ensures that the
pile has only one surface with the critical slope \cite{foot0}. The
width of the interface, in the steady state, can be measured in two
ways
\begin{eqnarray}
W_1^2(L) &=& \frac{1}{L}\sum_{i=1}^L\langle [h_i(t)-\langle
h_i\rangle]^2\rangle ~~~~~~\text{and,}\nonumber \\
W_2^2(L) &=& \frac{1}{L}\sum_{i=1}^L \langle [h_i(t)-s(t)\cdot i]^2\rangle\,,
\label{eq:width}
\end{eqnarray}
where $s(t)=2[\sum_{i=1}^L h_i(t)]/L(L+1)$ is mean slope of the
interface about which sum of {\it instantaneous} height fluctuations
vanish $\sum_{i=1}^L[h_i(t)-s(t)\cdot i]=0$.  In both definitions
(\ref{eq:width}) above the ensemble average $\langle\cdots\rangle$ is
identical to the time average in the steady state.
\begin{figure}[tb]
    \centering \includegraphics[angle=0,width=0.9\linewidth]{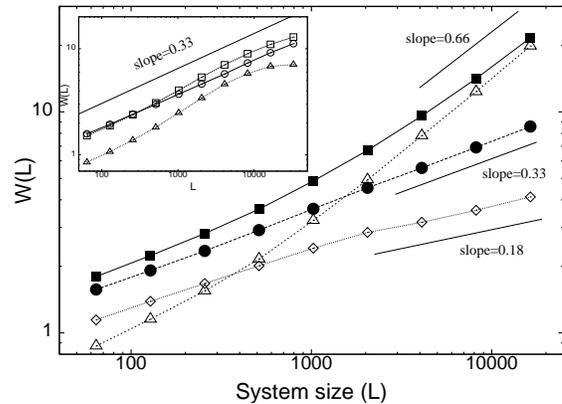}
    \caption{Width vs system size for LLS: $W_1$ (circles) and $W_2$
    (squares) are obtained using (\ref{eq:width}) while $W_S$
    (triangles) is computed from (\ref{eq:widthS}) below.  $W_c$
    (diamonds) $\sim L^{0.18}$ asymptotically. Inset: $W_1$ (circles)
    and $W_2$ (squares) and $W_S$ (triangles) for the ILLS with
    $r=20$.  }
\label{fig:w12.ps}
\end{figure}For large enough system sizes the scaling hypothesis is expected
to be valid and the roughness exponent $\alpha$ is defined through
\begin{equation}
W_{1,2}(L)\sim L^\alpha, ~~~L\rightarrow \infty \,.
\end{equation}
In Fig.~\ref{fig:w12.ps}, we plot the two widths $W_1$ and $W_2$ for a
range of system sizes $L=2^n$ with $6\leq n \leq 14$ (that is $64 \leq
L \leq 16384$).  It is seen that the two widths have different
asymptotic behaviours and give two different values of $\alpha$.  The
roughness $\alpha\simeq 0.33$ computed from $W_1$ (and $W_2$ up to
$L\lesssim L_c$) is in accordance with the predictions of
Eq.(\ref{eq:htevol}). The asymptotic roughness computed from $W_2$,
beyond the crossover system size $L_c\sim 1024$, $\alpha=0.65\pm 0.02$
is larger than that for an interface generated by a simple random
walk.  This is surprising since, {\it prima facie}, the
definitions in (\ref{eq:width}) are expected to be equivalent as far
as asymptotic scaling is concerned (\cite{foot001}). In fact, for the
ILLS, we have computed $W_1$ and $W_2$ for $r=20$
(Fig.~\ref{fig:w12.ps}, inset) and it is seen that they indeed have
the same asymptotic behaviour and hence, a unique value of
$\alpha\simeq 0.33$.

We now show that the crossover in $W_2$ in the LLS can be traced
to the fact that the time averaged steady state profile is not
linear, i.e., $\langle h_i(t)\rangle \neq \bar{s}~i$ and this
makes the two definitions nonequivalent. Although, this
nonlinearity of $\langle h_i(t)\rangle$ is a consequence of the
boundary effects and is present for the ILLS as well, we
demonstrate that the presence of SOC in LLS results in special
scaling properties which is responsible for the observed
differences between asymptotic scalings of $W_1$ and $W_2$.
It can be shown that $W_1$ and $W_2$ are related by
\begin{equation}
W_2^2(L)+W_c^2(L) = W_1^2(L)+W_S^2(L) \,,
\label{eq:Wdecomp}
\end{equation}
where $W_S(L)$ is the root-mean-square (rms) wandering of the steady state
interface profile $\langle h_i\rangle$,
\begin{equation}
W_S^2(L)\equiv \frac{1}{L} \sum_{i=0}^L \left[ \langle h_i\rangle -\bar
s\cdot i\right]^2 \,, \label{eq:widthS}
\end{equation}
with $\bar{s}=\langle s(t)\rangle$ being the mean slope of the steady
state profile satisfying $\sum_{i=0}^L [\langle h_i\rangle -\bar
s\cdot i]=0$.  The second term on the left of (\ref{eq:Wdecomp}),
$W_c^2(L)=\langle [s(t)-{\bar s}]^2\rangle(L+1)(2L+1)/6$, is the
contribution to interface width due to instantaneous slope
fluctuations and is the analogue of the {\it center of mass}
fluctuations in the case of moving interfaces \cite{krug97}. In fact,
in the case of moving interfaces $W_c$ dominates for large length
scales and hence $W_2$ is taken as the width. As
it turns out, for the anchored interfaces we deal with here, $W_c$ is
negligible compared to the other terms in (\ref{eq:Wdecomp}) as
$L\rightarrow \infty$.  From Fig.~\ref{fig:w12.ps}, it is evident
that the crossover in $W_2(L)$ for the LLS is due to the contribution
from $W_S$.  In the following we show that, for the LLS, the special
form of the nonlinear steady state profile $\langle h_i(t)\rangle$,
which results in $W_S \sim L^{0.66}$, is a consequence of the singular
diffusion associated with the SOC state.

In \cite{carlson90}, using a simple model, Carlson et. al.
demonstrated that the SOC state of the system is associated with the
vanishing of the density of {\it troughs} concommittant with the
divergence of the corresponding diffusion coefficient. The troughs are
defined as the sites for which $z_i\le 0$ so that, in the LLS, an
avalanche necessarily stops at a trough (or leave the system at
$i=0$). On a coarse grained level we define a set of densities $\{
\rho_n(x,t); n=-\infty,\cdots, -1,0,1,2\}$ where $\rho_n(x,t)$ denotes
the local density of sites with $z_i=n$. It follows that the coarse
grained local slope $z(x,t)$, which is locally conserved by the
dynamics, may be expressed in terms of the $\rho_n$'s as
$z(x,t)=\sum_{n=-\infty}^2 n\rho_n(x,t)$.  The open boundary condition
at $i=0$ implies $h(0,t)=0$ and the closed boundary condition at $i=L$
is modeled by setting $z(L,t)=0$. The trough density $\rho(x,t)\equiv
\sum_{n=-\infty}^0 \rho_n(x,t)$ is not strictly conserved. In the '01'
model considered in \cite{carlson90} the slope has only two values
$z_i=0$ (trough),$1$ and hence both $z$ as well as $\rho$ are strictly
conserved and are related simply as $z=1-\rho $. Although $\rho$ is
not strictly conserved for LLS and ILLS we still approximate its
dynamical evolution by the continuity equation, $\partial_t
\rho(x,t)+\partial_x J(\rho(x,t))=0$. Phenomenologically, the trough
current $J$ consists of three parts: (i) current due to addition of
grains $J_0$, (ii) avalanche current $J_A$, and (iii) a microscopic
noise term $\eta(x,t)$. In analogy with driven diffusive systems
\cite{krug92} the phenomenological form of the avalanche current is
written as $J_A(\rho)=a\rho+b\rho^2+\cdots-D(\rho)\partial_x\rho$.  As
in the 01 model, the critical state of the system
($L\rightarrow\infty,~r=\infty$) is associated with
$\rho,~\partial_x\rho \rightarrow 0$ and hence in order to
balance the finite input flux $J_0$, the diffusion constant
$D(\rho\rightarrow 0)$ has to diverge appropriately
\cite{carlson90}. Without loss of generality the leading divergence in
$D$ is taken to be a simple pole and the evolution of $\rho(x,t)$ is
thus \cite{foot01},
\begin{equation}
\frac{\partial\rho}{\partial t}=\frac{\partial}{\partial x}\left[
D(\rho)\frac{\partial\rho}{\partial x}+\eta\right];
~~D(\rho)=\frac{A(\rho)}{\rho^\phi} \,.
\label{eq:diff}
\end{equation}
Here $A(\rho)$ is taken to be a smoothly varying function in the
relevant interval $0\le\rho\le 1$.

\begin{figure}
    \centering
    \includegraphics[angle=0,width=0.8\linewidth]{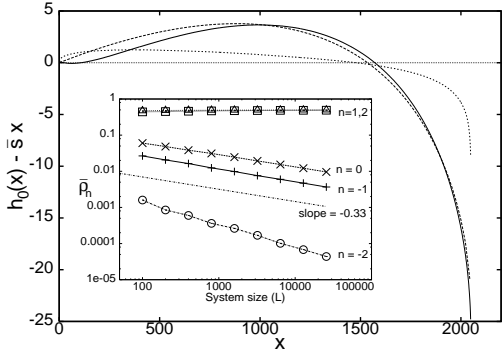}
    \caption{Deviation of the steady state height profile from
    straight line. For the LLS, the simulations data (solid curve) is
    well approximated by the analytic form (\ref{eq:heightprof})
    (dashed curve) for $L=2048$ with optimally chosen values of
    parameters $\rho_L$ and $\gamma$. The dotted curve is for the ILLS
    with $r=20$. Inset: Average density $\rhobar_n$ of sites with
    slope $z_i=n$ in the steady state as a function of system size
    $L$. }
\label{fig:fig3}
\end{figure}

In order to compute the steady state density profile
$\rho(x)=\rho(x,t\rightarrow\infty)$ of troughs, we note that, as
$L\rightarrow\infty$, the system is arbitrarily close to criticality
($\rho \rightarrow 0$), and hence $A(\rho)$ may be set to a constant
$A_0=A(0)$.  One can then readily integrate the current equation
$D(\rho)d\rho/dx=-J_0$, subjected to the boundary condition
$\rho(L)=\rho_L$, to obtain
\begin{equation}
\rho(x)\approx \rho_L\left[1+\gamma~(L-x)\right]^{-\theta}
\label{eq:ssprofile}
\end{equation}
where $\theta=1/(\phi-1)$ and $\gamma=-J_0~\rho_L^{1/\theta}/(\theta
A_0)$. The average density of the troughs $\rhobar\equiv L^{-1}\int_0^L
\rho(x) dx$ thus scales with system size asymptotically as $\bar\rho
\simeq \rho_L{(\gamma L)}^{-\theta}/(1-\theta)$.  From
Fig.~\ref{fig:fig3} (inset), we obtain $\theta\simeq 0.33$ and thus
$\phi\simeq 4$ for the LLS, which are the same as the corresponding
values obtained in \cite{carlson90} by direct measurement of $D(\rho)$
in a closed system.

In the following we find an approximate numerical relation between
$\rho(x)$ to $z(x)$ which would enable us to find the steady state
interface profile $h_0(x)=\int^x_0 z(x) dx$ form (\ref{eq:ssprofile}).
In Fig.~\ref{fig:fig3} (inset), we plot the average densities
$\rhobar_n \equiv L^{-1}\int_0^L dx \langle \rho_n(x,t)\rangle$ in the
steady state as functions of system size $L$. We note that, as $L$
increases, while $\rhobar_1 \simeq \rhobar_2\approx 1/2$ remain
finite, densities of all the trough sites vanish algebraically with
$L$: $\rhobar_0,\rhobar_{-1} \sim L^{-0.33}, \rhobar_{-2}\sim
L^{-0.6}$, and $\rhobar_n$'s with $n \le -3$ are negligible.  Hence,
in the limit of large $L$, we can approximate the total trough density
as $\rhobar\simeq \rhobar_0 +\rhobar_{-1}$. From the normalization
condition $\sum_{n=-\infty}^2 \rho_n(x,t) =1 $ it follows that
$\rho_1+\rho_2 = 1-\rho$ and we find numerically that
$\rhobar_1-\rhobar_2 \simeq 0.46 \rhobar$. Thus, for large systems, we
may write $ z(x) \simeq 2\rho_2 +\rho_1-\rho_{-1} \simeq
3/2-\kappa \rho(x)$, where numerically $\kappa\simeq 2$
\cite{foot1}.  Thus, $h_0(x)$ is given approximately by
\begin{equation}
h_0(x)\approx
\frac{3}{2}x-\frac{\kappa\rho_L(1+\gamma L)^{1-\theta}}{\gamma(1-\theta)}
\left(1-\left[\frac{1+\gamma(L-x)}{1+\gamma L}\right]^{1-\theta}\right).
\label{eq:heightprof}
\end{equation}
In Fig.~\ref{fig:fig3}, using the approximate expression for $\rho(x)$
from (\ref{eq:ssprofile}) with $\phi=4$, we compare $h_0(x)$ with that
obtained numerically and notice that the agreement is rather good
given the nature of approximations involved \cite{foot2}.

The $L$ dependence of $W_S$ of the steady state profile can be computed
from (\ref{eq:heightprof}) and turns out to be
\begin{equation}
W_S(L) \sim L^{1-\theta} \,.
\label{eq:widthS2}
\end{equation}
For the LLS with $\theta\simeq 0.33$, $W_S(L)\sim L^{0.66}$ dominates
$W_1\sim L^{1/3}$ and $W_c\sim L^{0.18}$ in (\ref{eq:Wdecomp}) and
hence, for large $L$, $W_2(L) \sim L^{0.66}$ -- in very good agreement
with our numerical results (Fig.~\ref{fig:w12.ps}). It is interesting
to note that $W_1$ and $W_2$ would have different asymptotic
behaviours if $\theta < 2/3$, i.e., if $\phi > 5/2$. For the `01'
model studied in \cite{carlson90} it is shown that $\phi=3$ exactly
and thus $W_2(L)\sim L^{1/2}$ for the corresponding interface
\cite{unpublished}.  Our conjecture that $W_1$ and $W_2$ would have
different asymptotic behaviour for SOC systems would be invalid if one
can devise a model with $\phi < 5/2$.

Next, we present a general argument as to why, for noncritical models
such as the ILLS, we do not expect any ambiguity in roughness
exponent. The essential difference between a critical and a
noncritical model is the presence of an additional length scale (apart
from system size and the microscopic cutoff) which shows up in the
avalanche size distribution (e.g. Fig. \ref{fig:sandpile} for ILLS) as
well as governs the decay of the boundary effects into the bulk.  In
the ILLS this length scale is related to the parameter $r$. To see
this we first note that as a cluster moves down its momentum $p$ makes
a random walk \cite{krug92}. If the mean density of troughs $\rho<1/r$
then most avalanches leave the system resulting in net drainage and if
$\rho>1/r$ then most avalanches stop on the pile leading to net growth
of the pile. Thus, in the steady state one has a finite density of
troughs $\rho=1/r$ and thus the mean spacing between the troughs
$1/\rho =r$ appears as the additional length scale. As already
discussed above, the current $J(\rho)$ now has a nonzero systematic
part $a\rho+b\rho^2\dots$, in addition to the finite diffusion
term. Such a systematic current, e.g., a term such as $a\rho$, results
in the effects of the boundaries decaying exponentially inside the
bulk. This is in contrast to the long ranged power law decay in the
LLS which is reflected in the forms of $\rho(x)$ and
$h_0(x)$. Although the steady state surface is nonlinear
(Fig.~\ref{fig:fig3}), its rms wandering $W_S$ is bounded and does not
alter the asymptotic scaling of $W_2$.

\begin{figure}[tb]
    \centering \includegraphics[angle=0,width=0.8\linewidth]{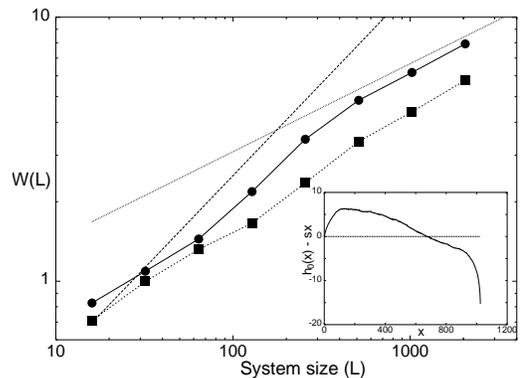}
    \caption{Interface width of the disordered sandpile as a function
    of system size. The squares represent $W_2$ and circles represent
    $W=\sqrt{W_2^2-W_S^2}$. The parameters chosen are the same as in
    \protect\cite{barker00}.  The straight lines have slope 0.33
    (dotted) and 0.72 (dashed). Inset: The steady state interface
    profile, after subtracting the mean linear profile.}
\label{fig:diswidth}
\end{figure}

Lastly, we briefly point out possible pitfalls of using $W_2$ naively
without properly accounting for the underlying steady state profile,
even in systems which do not show SOC. Recently, Barker and Mehta
\cite{barker00} studied a disordered version of the LLS where disorder
was introduced by allowing the added grains to
have an aspect ratio different from unity: grains are now
rectangles and are deposited on the sandpile with fixed probabilities
of landing on their larger or smaller edges. Thus, the height $h_i$ of
column $i$, no longer an integer, is the sum of the vertical edges of
all the grains in that column. In addition to the threshold dynamics
of LLS, dynamical reorientation of cluster of grains were allowed.
They found that the larger sandpiles cease to display SOC which is
reflected in the emergence of a preferred size of the large avalanches
in the associated drop number distributions.  Their numerical studies
of $W_2$ showed that while for very small systems ($L\lesssim 100$)
the roughness exponent $\alpha \simeq 0.34$, it seems to crossover to
a much larger value $\alpha \simeq 0.72$ for larger system sizes,
$100\lesssim L\lesssim 400$ (Fig.~\ref{fig:diswidth}).  Our preliminary
numerical studies, using the same parameters as in
\cite{barker00}, of yet larger systems ($500\lesssim L\le 2048$) show
that in fact the crossover seen in \cite{barker00} is evidently
transient and the asymptotic roughness exponent goes back to
$\alpha\simeq 0.33$ (see Fig.~\ref{fig:diswidth}).
 In fact, the crossover reported in \cite{barker00} is less noticeable
if one looks at $W=\sqrt{W_2^2-W_S^2}$ instead of $W_2$
(Fig.~\ref{fig:diswidth}) \cite{foot3}. This is in accordance with our
conjecture since the disordered system is not critical, although the
steady state profile is nonlinear (Fig.~\ref{fig:diswidth}, inset), it
does not affect the asymptotic roughness measured by $W_2$.  The
transient crossover seen in $W_2$ here is similar to what is seen for
the ILLS in Fig.~\ref{fig:w12.ps}, inset.

In summary, we have studied the surface roughness of a prototype
model of SOC and its modifications in one dimension. We find that
one needs to be careful in defining quantities such as the
interface width since special form of the steady state shape of
the surface in systems with SOC can result in different
asymptotic behaviours of otherwise equivalent definitions.
Although there is no such ambiguity for noncritical models, still
there may be crossovers at intermediate length scales which should
not be taken as the true asymptotic behaviour.

JGO and GT acknowledge financial support of {\it Centro de F\'{\i}sica do
Porto}, University of Porto where part of this work was
carried out. JGO also acknowledges financial support of {\it
Universidade de Aveiro}. JFFM was partially supported by project
POCTI/1999/FIS/33141 and POCTI/2002/MAT/46176.

\end{multicols}

\begin{references}
\bibitem{foot001} However, as shown in (\ref{eq:Wdecomp}), $W_1$ and
$W_2$ are not numerically identical. E.g. for an anchored but flat
surface which fluctuates as a whole $W_2=0$ but $W_1\ne 0$ whereas for
a static non-planar surface $W_1=0$ but $W_2\ne 0$.
\bibitem{foot01} One could write the evolution equation for $z(x,t)$
directly as $\partial_t z(x,t)+\partial_x J(z,x,t)=0$, with the
diffusion constant diverging as at $z=3/2$ for the LLS. However, for
different models the critcal slope would be different form $3/2$ and
hence it is more natural to write the equation for $\rho$ with the
singularity always at $\rho=0$.
\bibitem{foot1} We have numerically checked that, except for points
close to the boundaries $x\simeq 0,~L$, similar approximations hold
good for the local densities $\rho_n(x,t)$ as for the spatially
averaged ones.
\bibitem{foot2} The deviations seem most pronounced near the left
boundary $x=0$ which can be improved by including more terms in
$A(\rho)=A_0+A_1\rho+\dots$.
\bibitem{foot3} In the disordered case the slope fluctuations are
large and hence $W_1$ is dominated by $W_c$ and is numerically larger
than $W_2$ in the range of system sizes we study.
\end{references}
\end{document}